\documentclass{aa}
\usepackage{aabib}
\usepackage{graphicx}
\def\aj{AJ}			
\def\araa{ARA\&A}		
\def\apj{ApJ}			
\def\apjl{ApJ}		
\def\apjs{ApJS}

\def\aap{A\&A}		
		
\def\aaps{A\&AS}

\def\mnras{MNRAS}

\def\pasp{PASP}		
\def\pasj{PASJ}

\def\ssr{Space~Sci.~Rev.}

\begin{document}
   \title{Chandra observation of an unusually long and intense X-ray flare 
from a young solar-like star in M\,78}

\titlerunning{An unusually long and intense X-ray flare from a young solar-like star in M\,78}
\authorrunning{Grosso et al.}

%  \subtitle{}

   \author{N. Grosso\inst{1}$^\mathrm{,}$\inst{2}
      \and T. Montmerle\inst{2}
      \and E.D. Feigelson\inst{3}
      \and T.G. Forbes\inst{4}
          }

   \offprints{Nicolas Grosso, at the current email address~: Nicolas.Grosso@obs.ujf-grenoble.fr\,.}

   \institute{Max-Planck-Institut f{\"u}r extraterrestrische Physik,
              P.O. Box 1312, D-85741 Garching bei M{\"u}nchen, Germany
              \and
              Laboratoire d'Astrophysique de Grenoble, 
              Universit{\'e} Joseph-Fourier,
              F-38041 Grenoble cedex 9, France
         \and Department of Astronomy and Astrophysics, 
              Pennsylvania State University, 525 Davey Laboratory, 
              University Park, PA 16802, USA
	 \and Institute for the Study of Earth, Oceans, and Space, 
	      39 College Road, University of New Hampshire, 
	      Durham, NH 03824, USA
             }

   \date{Received 7 July 2003 / Accepted 29 January 2004}

   \abstract{\object{LkH$\alpha$\,312} has been observed serendipitously with the ACIS-I detector 
on board the {\sl Chandra X-ray Observatory} with 26\,h continuous exposure. This H$\alpha$ 
emission line star belongs to \object{M\,78} (\object{NGC\,2068}), one of the star-forming regions 
of the Orion~B giant molecular cloud at a distance of 400\,pc. From the optical and the near-infrared 
(NIR) data, we show that LkH$\alpha$\,312 is a pre-main sequence (PMS) low-mass star with a weak NIR excess.
This genuine T~Tauri star displayed an X-ray flare with an unusual long rise phase ($\sim$8\,h). 
The X-ray emission was nearly constant during the first 18\,h of the observation, and 
then increased by a factor of 13 during a fast rise phase ($\sim$2\,h), and reached a factor of 16 
above the quiescent X-ray level at the end of a gradual phase ($\sim$6\,h) showing a slower rise. 
To our knowledge this flare, with $\sim$0.4--$\sim$0.5\,cts\,s$^{-1}$, has the highest count rate 
observed so far with {\sl Chandra} from a PMS low-mass star. 
By chance, the source position, $8.2\arcmin$ off-axis, protected this observation from pile-up. 
We make a spectral analysis of the X-ray emission versus time, showing that the plasma temperature of the quiescent phase and the 
flare peak reaches 29\,MK and 88\,MK, respectively.
The quiescent and flare luminosities in the energy range 0.5--8\,keV corrected from absorption 
($N_{\rm H} \approx 1.7\,10^{21}$\,cm$^{-2}$) are $6\,10^{30}$\,erg\,s$^{-1}$ and 
$\sim$10$^{32}$\,erg\,s$^{-1}$, respectively. 
The ratio of the quiescent X-ray luminosity on the LkH$\alpha$\,312 bolometric luminosity is very high with 
$\log (L_{\rm X}/L_{\rm bol})= -2.9$, implying that the corona 
of LkH$\alpha$\,312 reached the `saturation' level. The X-ray luminosity of the flare peak reaches $\sim$2\% 
of the stellar bolometric luminosity. 
The different phases of this flare are finally discussed in the framework of solar flares, 
which leads to the magnetic loop height from 3.1$\,10^{10}$ to $10^{11}$\,cm (0.2-0.5\,R$_\star$, i.e., 0.5--1.3\,R$_\odot$).
               }

\maketitle

\begin{figure*}[!t]
\centering
\includegraphics[angle=90,width=18cm]{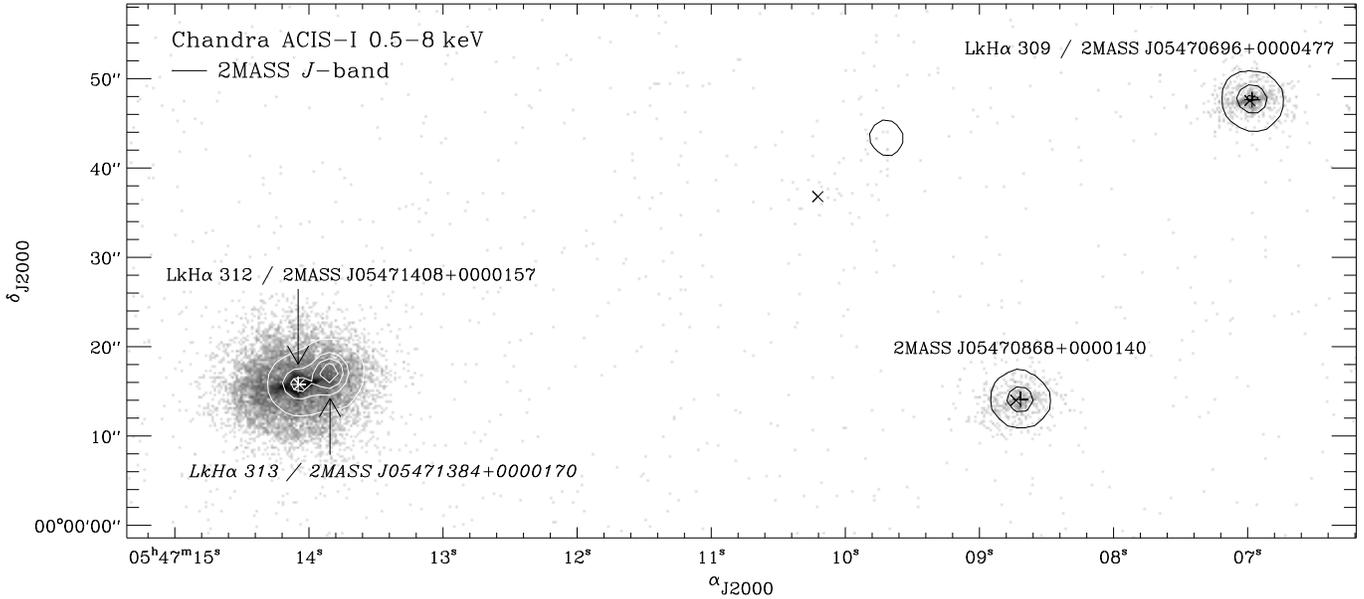}
 \caption{{\sl Chandra} observation of LkH$\alpha$\,312 on October 18, 2000. The background image is 
the ACIS-I observation in the energy range 0.5--8\,keV of LkH$\alpha$\,312, 
located at $\sim8\arcmin$ off-axis (the pixel size is 0\farcs25). 
The color scale is stretched logarithmically to enhance the point spread function (PSF) features, 
note the $2\arcsec$ asymmetric cusp surrounded by a $10\arcsec$ elliptical halo. 
The contour map displays arbitrary intensities from the 2MASS $J$-band image 
using linear scale. The positions of X-ray sources detected above 
the threshold of $3\,10^{-7}$ are marked with `$\times$'. 
Their infrared positions in the 2MASS survey are marked by `+'.
}
\label{map}
\end{figure*}

\section{Introduction}

Since the first X-ray observations of star-forming regions (SFR) with the {\sl Einstein Observatory}, 
young PMS low-mass stars, T~Tauri stars (TTS), are known to be variable in X-rays 
(\cite{feigelson81}; \cite{montmerle83}). 
TTS, as active stars, display X-ray flares triggered by magnetic reconnection events 
occurring in their stellar coronae (see reviews by \cite{feigelson99}, and \cite{favata03}). 
The {\sl Chandra X-ray Observatory}, thanks to its highly elliptical orbit that permits 
continuous observation over many hours, has collected a real zoo of X-ray flares from TTS in 
several SFR (e.g., \cite{preibisch02}, \cite{feigelson02a}, \cite{imanishi03}). The X-ray flares 
of TTS are generally impulsive, displaying a fast rise phase ($\sim$2\,h) corresponding to a heating 
phase, followed by an exponential decay corresponding to a cooling phase (e.g., \cite{imanishi03}). 
Only a few long duration flares have been observed from TTS with the previous generation of X-ray 
satellites, {\sl ROSAT} and {\sl ASCA} (\object{SR13}, \cite{casanova94}; object{V773\,Tau}, \cite{skinner97}; 
\object{P1724}, \cite{stelzer99}). \cite*{stelzer99} showed that these light curves can be reproduced 
by the rotational modulation of the exponential decay of a cooling flare.  
\cite*{feigelson02a}, investigating the variability of young solar-like stars 
(M$_\star$=0.7--1.4\,M$_\odot$) in the Orion Nebula Cloud, find several likely long-duration flares 
but the low count rates of these events prevent a time-dependent spectroscopy analysis.

M\,78 (NGC\,2068) is a reflection nebula illuminated by a B1.5V star (HD\,38563 North; \cite{mannion84}), 
and located in the northern part of the closest giant molecular cloud, Orion~B (L\,1630), 
at a distance of $\sim$400\,pc (\cite{anthony82}). 
The star-forming region M\,78 has been observed with the ACIS-I detector 
aboard {\sl Chandra} with 26\,h exposure on October 18, 2000 
(sequence number 200100). We report here the study of the brightest X-ray 
source detected during this observation, LkH$\alpha$\,312, located $\sim10\arcmin$ at the 
South-East of the optical emission nebula M\,78. This source displays an intense X-ray flare with 
an unusually long rise phase. 
We describe the ACIS-I data reduction and the X-ray detection of LkH$\alpha$\,312 
in \S\ref{observation}. The evolutionary status of LkH$\alpha$\,312 is determined from 
the optical and the NIR data in \S\ref{status}. The {\sl Chandra} light curve of the flare 
is presented in \S\ref{variability}. The time-dependent spectra is investigated in 
\S\ref{spectroscopy}. Finally, the different phases of this flare are discussed in the framework 
of solar flares in \S\ref{discussion}. Concluding remarks are presented in \S\ref{conclusion}.

\section{Chandra ACIS-I observation}
\label{observation}

The telemetry format of this observation is `Timed Exposure' with 3.2\,s for the frame time,
combined with the ACIS mode `Faint'. The aimpoint of the ACIS-I detector is 
$\alpha_\mathrm{J2000}=05^\mathrm{h}46^\mathrm{m}44\fs2$,  
$\delta_\mathrm{J2000}=+00\degr03\arcmin46\farcs0$ 
($l=205.3\degr$, $b=-14.3\degr$), centered on the optical emission nebula M\,78. 

\begin{table}[!ht]
\caption{{\sl Chandra} X-ray sources with 2MASS counterpart in 
the vicinity of LkH$\alpha$\,312 and their X-ray count rates on October 18, 2000.}
\label{cxc}
\begin{tabular}{@{}crccc@{}}
\hline
\hline
\noalign{\smallskip}
CXOU 0547    & \multicolumn{1}{c}{Count rate$^\mathrm{a}$} & 2MASS\,J0547      & r$^\mathrm{b}$      & LkH$\alpha$ \\
            & [cts\,ks$^{-1}$]  &                  & [$\arcsec$] &\\
\noalign{\smallskip}
\hline
\noalign{\smallskip}
06.9+000047 &   6.8$\pm$0.3  & 0696+0000477       & 0.3    & 309   \\
08.7+000013 &   5.7$\pm$0.3  & 0868+0000140       & 0.5    & \dotfill      \\
14.0+000015  & 149.8$\pm$1.3  & 1408+0000157       & 0.1    & 312   \\     
\noalign{\smallskip}
\hline
\end{tabular}
\begin{list}{}{}
\small
\item[$^{\mathrm{a}}$] In the energy range 0.5--8\,keV, from {\tt wavdetect} taking 93859\,s for the exposure.
\item[$^{\mathrm{b}}$] Offsets between X-ray and NIR positions.
\end{list}
\end{table}

Following \cite*{feigelson02b} we start our data analysis from the level~1 
processed event list provided by the {\sl Chandra X-ray Center} (CXC) with the pipeline 
processing version R4CU5UPD11.
We first use the tool {\tt acis\_process\_events} of 
the {\sl Chandra Interactive Analysis of Observations} ({\tt CIAO}) package 
(version 2.2.1 and 2.3\footnote{\tt http://asc.harvard.edu/ciao}) to remove the unnecessary 
$\pm0.25\arcsec$ randomization introduced by the pipeline processing on each event 
position, and we select events belonging to Good Time Intervals (GTI) provided by the CXC.
The observation time corrected from deadtime (after each CCD exposure 0.04\,s are lost 
to read the CCD chip) is 93859\,s, i.e., more than 26\,h. 
Second we correct the energy and grade of each event using the Penn State's 
Charge Transfer Inefficiency (CTI) corrector package
(version 1.41\footnote{\tt http://www.astro.psu.edu/users/townsley/cti/\-install.html}; 
\cite{townsley00}; \cite{townsley02}), which models the CTI characteristics of each CCD chip.
The most fundamental result of this CTI correction is the improvement in spectral resolution.
Third we clean the data by selecting only events with ASCA grade (0, 2, 3, 4, and 6) 
and energy in the range 0.5--8\,keV after CTI correction. 
Bad pixels and hot columns are rejected by selecting only events with 
status bits 1--15 and 20--32 equal to zero. 
Remaining events with non zero status bits are events which were identified by the pipeline 
processing as produced by cosmic ray afterglows. However few of them are obvious events 
from bright X-ray sources misclassified by the pipeline. Hence we reject these events
only during X-ray source detection to avoid spurious X-ray sources, but we use them 
for source spectral analysis. 
Finally to improve the absolute event positional accuracy, 
we correct the aspect offset using the fix offsets 
thread\footnote{\tt http://cxc.harvard.edu/cal/ASPECT/fix\_offset/\-fix\_offset.cgi}.  

Fig.~\ref{map} presents an enlargement of the resulting X-ray image at an off-axis 
angle $\theta\sim8\arcmin$.
Three X-ray sources counterparts of optical emission-line stars (\cite{herbig63})
and/or NIR 2MASS sources (all-sky data release; \cite{cutri03}) 
are visible. To obtain accurate X-ray positions and to detect weaker 
X-ray sources in this area, we use the tool {\tt wavdetect} of the {\tt CIAO} package, 
with the threshold probability of $3\,10^{-7}$, and 11 wavelet scales ranging 
from $0.5\arcsec$ to $16\arcsec$ in power of $\sqrt{2}$.
Table~\ref{cxc} gives the list of the {\sl Chandra} X-ray sources in Fig.~\ref{map} identified 
with 2MASS counterparts. We find that the agreement between the X-ray and the NIR positions 
is better than 0\farcs5. 
No X-ray source is detected at the position of \object{LkH$\alpha$\,313}, 
but $4\arcsec$ away we find the remarkable X-ray detection of \object{LkH$\alpha$\,312} 
with more than 14,000 counts, i.e., an average count rate of $\sim$0.15\,cts\,s$^{-1}$. 

\begin{table}[!t]
\caption{{\sl ROSAT/HRI} observation of LkH$\alpha$\,312 from September 5 to October 5, 1997.}
\label{1rxh}
\begin{tabular}{@{}ccccc@{}}
\hline
\hline
\noalign{\smallskip}
ROR       & Exp.    & RX~J0547 & $\cal L^\mathrm{a}$ & CR$^\mathrm{b}$                 \\
2025      & [ks]    &          &            &  [cts\,ks$^{-1}$]        \\
\noalign{\smallskip}
\hline
\noalign{\smallskip}
21h-1 & 26.7  & 14.1+000011 & 213 & 8.3$\pm$0.7   \\ 
\noalign{\smallskip}
\hline
\end{tabular}
\begin{list}{}{}
\small
\item[$^{\mathrm{a}}$] $\cal L$ is the likelihood of existence.
\item[$^{\mathrm{b}}$] The count rate is for the {\sl ROSAT} energy band, 0.1--2.4\,keV.
\end{list}
\end{table}

\begin{figure}[!t]
\centering
\includegraphics[angle=90,width=8.8cm]{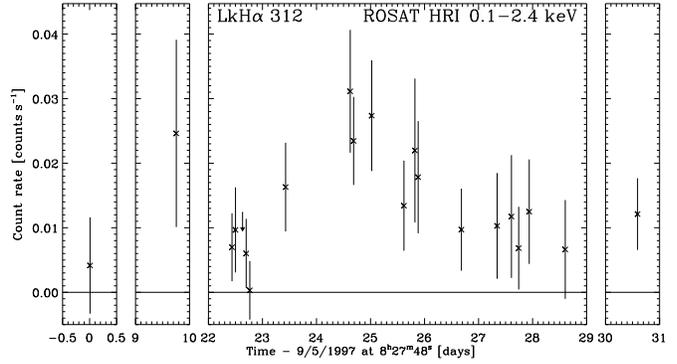}
 \caption{{\sl ROSAT/HRI} X-ray light curve of LkH$\alpha$\,312 obtained from September 5 to October 5, 1997.
Crosses show the X-ray light curve and one-sigma error bars 
in the energy range 0.1-2.4\,keV for each observational segment ($\sim$20\,min).
Background count rates are shown as upper limits when the background subtracted count rate 
is negative.
}
\label{rosat_lightcurve}
\end{figure}

We find a previous X-ray detection of this source made from September 5 to October 5, 1997, with the 
{\sl High Resolution Imager} (HRI) on board {\sl ROSAT} in the energy range 0.1-2.4\,keV. 
Table~\ref{1rxh} gives the results of our analysis of the archived 
{\sl ROSAT Observation Request} (ROR) with 
the {\tt EXSAS}\footnote{\tt http://wave.xray.mpe.mpg.de/exsas} package (\cite{zimmermann97}). 
The total exposure of 26.7\,ks is the sum of several $\sim20$\,min segment observations 
spread over one month. We extract source events from the 95\% encircled energy radius and produce 
a background subtracted light curve (corrected from the PSF fraction) 
with one bintime per segment (Fig.~\ref{rosat_lightcurve}).
As the {\sl HRI} has no spectral resolution, we estimate the {\sl ROSAT/HRI} count rate of 
the quiescent level with W3PIMMS\footnote{\tt http://asc.harvard.edu/toolkit/pimms.jsp} 
using the {\sl Chandra} light curve and spectroscopy results (see low \S\ref{variability} 
and \S\ref{spectroscopy}), which give a {\sl Chandra} quiescent count rate of 0.035\,cts\,s$^{-1}$, 
and a {\sl Chandra} spectrum model with $N_{\rm H}=1.73\,10^{21}$\,cm$^{-2}$, $Z\sim0.2\,Z_\odot$, 
$<kT>$=2.5\,keV. We find 0.004\,cts\,s$^{-1}$, which is consistent with the low level detected visible 
in the light curve. This source is clearly variable, however the poor sampling of the light curve 
precludes obtaining its detailed shape.

\section{Evolutionary status of LkH$\alpha$\,312}
\label{status}

LkH$\alpha$\,312 was originally selected due to its (weak) H$\alpha$ line emission in slitless grating survey 
of M\,78 (\cite{herbig63}). An optical spectrum with 7\,{\AA} resolution 
has been obtained in 1977 by \cite*{cohen79}, leading to its M0 spectral type classification, 
and its H$\alpha$ equivalent width, EW(H$\alpha$)=8.2\,{\AA} (see Table~\ref{data}). 
We note that this star is not classified as a TTS in the catalog of \cite*{herbig88},  
probably due to its low EW(H$\alpha$) value. 
\cite*{martin97} considers that classical TTS (CTTS) should have EW(H$\alpha$) 
larger than expected from the average chromospheric activity for their spectral type, 
and adopts EW(H$\alpha$)=10\,{\AA} for early-M stars as a lower limit, instead of the 
classical EW(H$\alpha$)=5\,{\AA} limit.
On the basis only of its EW(H$\alpha$), LkH$\alpha$\,312 could then be either an non-accreting PMS star 
of M\,78 or a foreground main-sequence dwarf with H$\alpha$ emission, a dMe star. 
Of course the detection of the Li\,{\small I} 6707 absorption line would definitively 
demonstrate the youthfulness of this star, but this observation is not available.

\begin{table*}[!t]
\caption{Optical and available near-infrared data for the emission line stars discussed in this article. See corresponding $J-H$, $H-K_{\rm S}$ diagram in Fig.~\ref{hr_diagram}, spectral energy distributions in Fig.~\ref{sed},  and H.-R.\ diagram in Fig.~\ref{colour_colour}.}
\label{data}
\begin{tabular}{@{}clcrcrrrrrcc@{}}
\hline
\hline
\noalign{\smallskip}
LkH$\alpha$ & \multicolumn{4}{c}{Cohen \& Kuhi (1979)} & \multicolumn{5}{c}{2MASS} & \multicolumn{2}{c}{This work} \\
\vspace{-0.6cm}\\
             &     \multicolumn{4}{c}{\hrulefill} & \multicolumn{5}{c}{\hrulefill} & \multicolumn{2}{c}{\hrulefill}\\
 & Spec. &  T$_{\rm eff}$ &  \multicolumn{1}{c}{H$\alpha$} & \multicolumn{1}{c}{$V$} & \multicolumn{1}{c}{$J^\mathrm{a}$} & \multicolumn{1}{c}{$H^\mathrm{a}$} & \multicolumn{1}{c}{$K_\mathrm{S}$$^\mathrm{a}$} & \multicolumn{1}{c}{$J$-$H$} & \multicolumn{1}{c}{$H$-$K_\mathrm{S}$} & A$_\mathrm{V}$ & L$_\mathrm{bol}$$^\mathrm{b}$  \\
\#     & type  &      [K]                                 & \multicolumn{1}{c}{[$\mathrm{\AA}$]}  &               \multicolumn{1}{c}{[mag]}             & \multicolumn{1}{c}{[mag]} & \multicolumn{1}{c}{[mag]} & \multicolumn{1}{c}{[mag]} & \multicolumn{1}{c}{[mag]} & \multicolumn{1}{c}{[mag]} &     [mag]   & [L$_\odot$]         \\
\noalign{\smallskip}
\hline
\noalign{\smallskip}
309$^\star$ & K7   & 4000 & 51.2  & 15.1   & 11.35$\pm$0.04 & 10.57$\pm$0.04 & 10.22$\pm$0.04 & 0.78$\pm$0.08 & 0.35$\pm$0.08 & 2.0$\pm$0.5  & 1.6$\pm$0.2 \\
312$~\,$ & M0      & 3917 &  8.2  & 15.1 & 11.32$\pm$0.05 & 10.52$\pm$0.06 & 10.12$\pm$0.06 & 0.80$\pm$0.11 & 0.40$\pm$0.12   & 1.8$\pm$0.5  & 1.5$\pm$0.3  \\
313$^\star$ & M0.5 & 3802 & 19.6  & 15.9   & 10.75$\pm$0.09 & 9.79$\pm$0.09 & 9.23$\pm$0.07 & 0.95$\pm$0.18 & 0.57$\pm$0.16   & 3.5$\pm$0.5  & 3.9$\pm$0.5  \\
\noalign{\smallskip}
\hline
\end{tabular}
\begin{list}{}{}
\small
\item[$^{\mathrm{a}}$] The errors are the total magnitude uncertainties given at the 90\% confidence level.
\item[$^{\mathrm{b}}$] From fits of Fig.~\ref{sed} and taking $d=400$\,pc.
\item[$^\star$] Classified as classical T~Tauri star in Herbig \& Bell (1988).
\end{list}
\end{table*}

\begin{figure}[!t]
\centering
\includegraphics[angle=0,width=7cm]{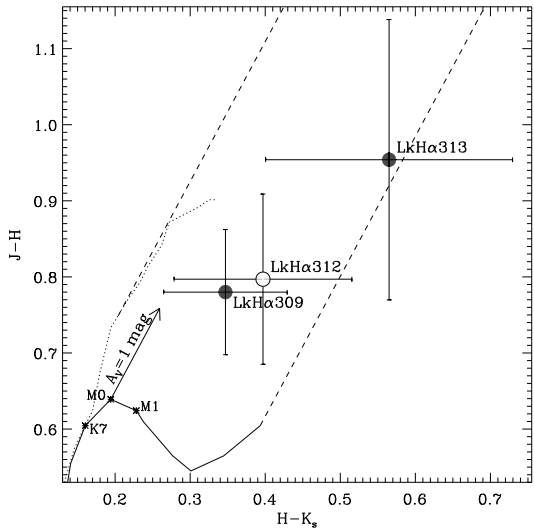}
 \caption{$J-H$, $H-K_\mathrm{S}$ diagram of LkH$\alpha$\,312 (open dot). 
Magnitude error bars are the total magnitude uncertainties at the 90\% confidence level.
The intrinsic colors of giants (dotted line) and A0--M6 dwarfs 
(continuous line) from Bessel \& Brett (1988), adapted for the 2MASS photometric system 
using the 2MASS color transformations (Carpenter 2001; Cutri et al.\ 2003), are plotted for comparison. 
The extinction law of Cohen et al.\ (1981), also adapted for the 2MASS photometric system, 
is drawn for giants and M9 stars. The arrow indicates a visual extinction of 1\,mag. 
Black dots mark the position of known classical T~Tauri stars in the vicinity of LkH$\alpha$\,312 (Herbig \& Bell 1988). 
}
\label{colour_colour}
\end{figure}

\begin{figure}[!t]
\centering
\includegraphics[angle=0,width=\columnwidth]{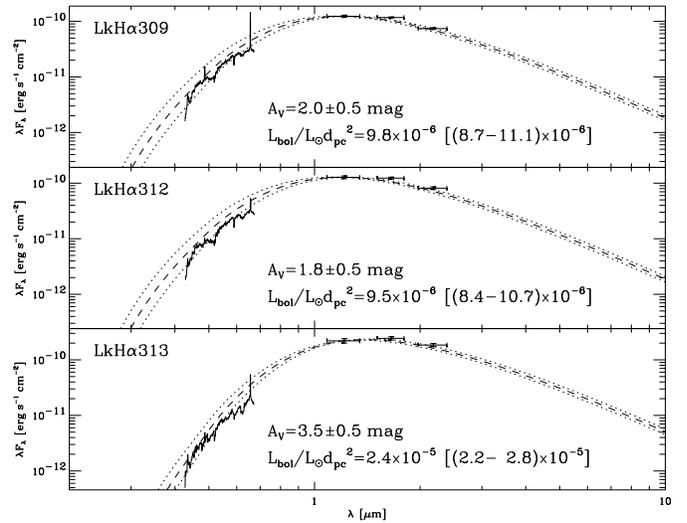}
 \caption{Spectral energy distributions. The optical spectra and the NIR photometry come 
from Cohen \& Kuhi (1979) and 2MASS, respectively.
Dashed line shows the absorbed black body spectrum having the effective temperature found by (Cohen \& Kuhi 1979), 
coinciding with the $J$-band photometry, and consistent with the optical spectra. Upper and lower dotted lines show the same 
black body spectrum for minus and plus 0.5\,mag extinction,respectively. We give the resulting bolometric luminosities, 
including typical ranges for a typical extinction error of 0.5\,mag.
}
\label{sed}
\end{figure}

Recently \cite*{preibisch02} have shown in the young star cluster IC\,348 
that H$\alpha$ emission is a poor indicator of circumstellar material around TTS.
More precisely, an IR excess in the $L$-band or beyond may be present at the same time as weak H$\alpha$ emission, indicating that such objects are in fact surrounded by a hollow circumstellar disk but currently are accretion quiet. 
A similar conclusion was drawn by \cite*{andre94} based on a sample of $\rho$ Ophiuchi dark cloud's TTS.
\cite*{preibisch02} use $K-L$ to trace the infrared excess, but $L$-band photometry is 
not available for LkH$\alpha$\,312. We have found in the ISO archives\footnote{\tt http://www.iso.vilspa.esa.es}, 
6.7\,$\mu$m and 14.3\,$\mu$m ISOCAM detections of the TTS LkH$\alpha$\,313, 
but the poor angular resolution of ISOCAM ($\sim$6\arcsec) cannot resolve the emission of 
LkH$\alpha$\,312 only $4\arcsec$ away.  
Fig.~\ref{colour_colour} displays in a $J-H$, $H-K_\mathrm{S}$ diagram the position of two known TTS 
of M\,78 selected due to their strong H$\alpha$ line emission in the vicinity of LkH$\alpha$\,312 (see Table~\ref{data}). 
Both TTS have a visual extinction, $A_{\rm V}$, greater than 1\,mag.
The NIR colors of LkH$\alpha$\,312 are very similar to those of the accreting TTS LkH$\alpha$\,309, 
and can be retrieved from the intrinsic color of a M0 star by applying $\sim$1.8\,mag of 
visual extinction combined with a small NIR excess, which is maybe produced by a circumstellar 
disk with an inner hole (\cite{andre94}). 
This extinction is consistent with LkH$\alpha$\,312 being a young star of M\,78 
rather than a foreground main-sequence dwarf.

We note that the extinction derived from the color-color diagram for LkH$\alpha$\,312 is a factor of $\sim$4 
larger than the one found by \cite*{cohen79} from optical spectroscopy. The extinction values obtained by \cite*{cohen79} 
for LkH$\alpha$\,309 and LkH$\alpha$\,313 are also smaller than the one measured in our color-color diagram. Moreover the bolometric 
luminosities, $L_{\rm bol}$, given by \cite*{cohen79} for these objects ``represents solely the optical data (extrapolated 
[with a black body] to infinite wavelength from 0.67\,$\mu$m)'' (see their Table~5, note of columm 17). To investigate this 
extinction discrepancy and to udpate the bolometric luminosities, we build spectral energy distributions (SED; Fig.~\ref{sed}). 
The optical spectra were copied directly from the Fig.~23 of \cite*{cohen79}. The 2MASS photometry was first transformed to the UKIRT 
system (Carpenter 2001; Cutri et al.\ 2003), and then converted to flux densities (\cite{cox00}).
We compute the black body spectrum having the effective temperature found by \cite*{cohen79}, we apply using \cite*{fitzpatrick99} 
the extinction given by \cite*{cohen79}, and we scale the absorbed black body spectrum for coincidence with the 0.67\,$\mu$m photometry. The obtained SED are far below 
the NIR data for all the stars of Table~\ref{data}, hence both the extinction and the bolometric luminosity of \cite*{cohen79} are underestimated, 
and need to be udpated. Taking the value found by \cite*{cohen79} for the effective temperature, we adjust the extinction to obtain an absorbed black body SED, coinciding with the 2MASS photometry in the $J$-band (the contribution of infrared-excess, if any, is negligible at this wavelength), and consistent with the optical spectrum. 
Fig.~\ref{sed} displays our best absorbed black body, with the resulting $L_{\rm bol}$ estimate. We consider a typical $A_{\rm V}$ error of 0.5\,mag. 
We list in Table~\ref{data} our values of $A_{\rm V}$ and $L_{\rm bol}$ for LkH$\alpha$\,309, LkH$\alpha$\,312, and LkH$\alpha$\,313.
The estimated extinctions are now consistent with the ones derived from the color-color diagram.

\begin{figure}[!t]
\centering
\includegraphics[angle=0,width=\columnwidth]{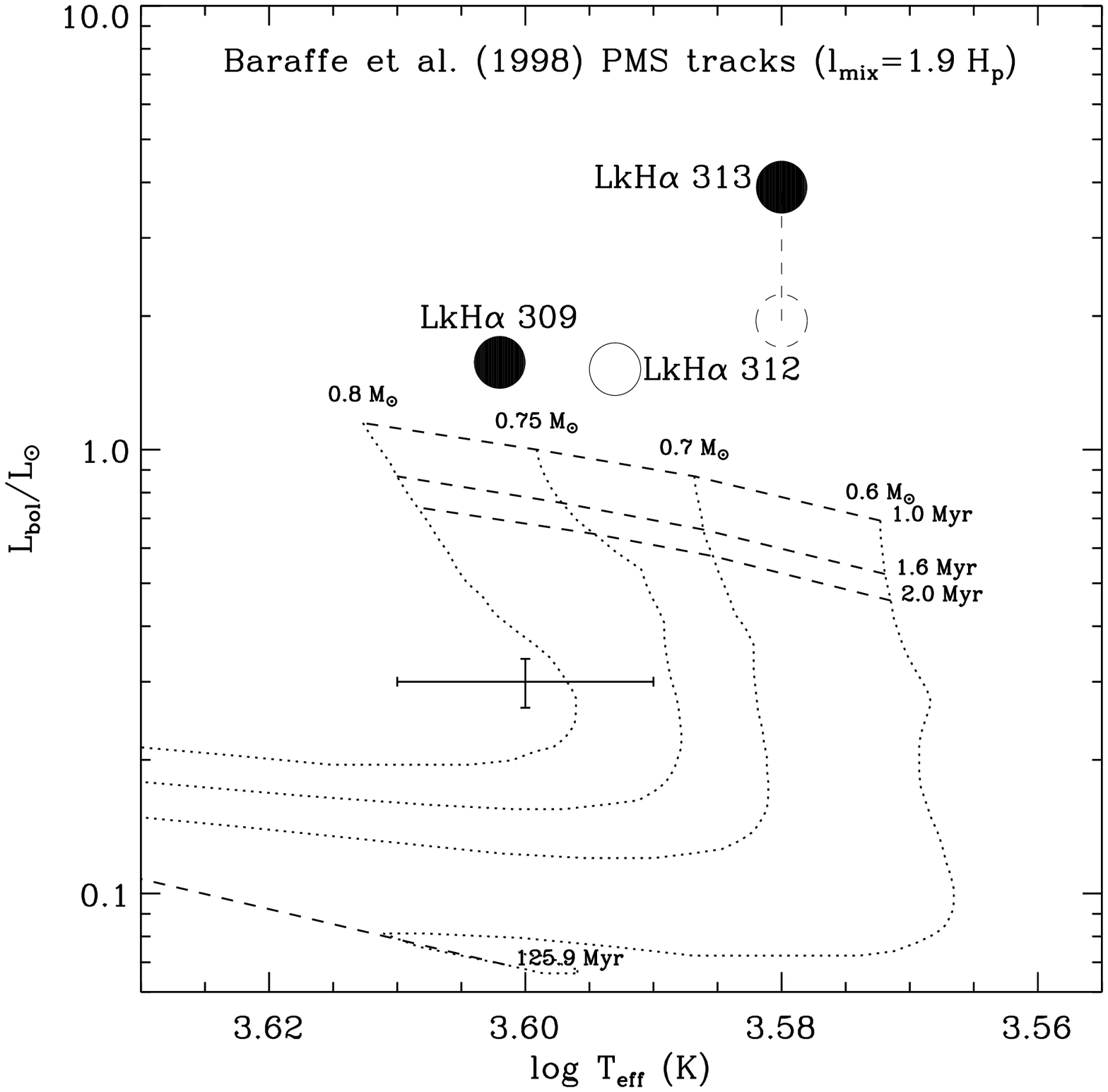}
 \caption{H.-R. diagram of LkH$\alpha$\,312 (open dot). 
Dotted lines show Baraffe et al.\ (1998) PMS tracks (with the mixing length required to fit the Sun, 
$l_\mathrm{mix}=1.9\,H_\mathrm{p}$, with $H_\mathrm{p}$ the pressure scale height). 
Dashed lines show the corresponding isochrones. Black dots mark the position of known classical 
T~Tauri stars in the vicinity of LkH$\alpha$\,312 (Herbig \& Bell 1988). The dashed circle marks the position of LkH$\alpha$\,313 in case of binarity. 
Error bar shows the typical error on the effective temperature (Cohen \& Kuhi 1979), 
and on the bolometric luminosity (see Fig.~\ref{sed}).
}
\label{hr_diagram}
\end{figure}

Fig.~\ref{hr_diagram} shows the H.-R. diagram with the PMS tracks of \cite*{baraffe98} 
fitting the Sun (i.e., having the mixing length parameter, $l_\mathrm{mix}$, 
equal to 1.9 times the pressure scale height, $H_\mathrm{p}$), with 
the position of LkH$\alpha$\,312 assuming that LkH$\alpha$\,312 and the other stars of 
Table~\ref{data} are at the distance of M\,78 ($d \sim 400$\,pc).
Considering LkH$\alpha$\,312 as a main sequence star would put it on 
the 120\,Myr isochrone revising down its distance to only $d\sim90$\,pc. 
This would locate LkH$\alpha$\,312 in the Local Bubble where  
$N_\mathrm{H} \sim 5\, 10^{19}$\,cm$^{-2}$ (\cite{sfeir99}).
This small distance would not be consistent with the extinction of this object, 
$A_\mathrm{V} \sim 1.8$\,mag, 
which implies $N_\mathrm{H} \sim 3.2\, 10^{21}$\,cm$^{-2}$ 
( using, e.g., the $N_\mathrm{H}/A_\mathrm{V}$ conversion factor of \cite{predehl95}). 
Therefore we consider LkH$\alpha$\,312 as a PMS star of M\,78.

From the \cite*{baraffe98} PMS tracks we find that
LkH$\alpha$\,312 is located above the 1\,Myr isochrone, among known CTTS of M\,78, 
with a stellar mass of 0.7--0.75\,M$_\odot$.
We also note that if LkH$\alpha$\,313 were a close binary system the luminosity of the individual 
components would be decreased by a factor of $\sim$2 (see Fig.~\ref{hr_diagram}). In 
this case LkH$\alpha$\,309, LkH$\alpha$\,312, and LkH$\alpha$\,313 would all be coeval, and LkH$\alpha$\,312 and LkH$\alpha$\,313 would form 
a gravitationally bound system since their separation is only 4\arcsec~on the sky.

Our final conclusion on the evolutionary status of LkH$\alpha$\,312 is that it is a {\it genuine} TTS, 
with maybe a transition circumstellar disk, belonging to the M\,78 star-forming region.
Combining $T_{\rm eff}=3917$\,K and $L_{\rm bol}=1.5$\,L$_\odot$ leads to the stellar 
radius $R_\star=2.6$\,R$_\odot$.

\section{Variability study}
\label{variability}

\subsection{Chandra X-ray light curve}

\begin{figure*}[!hp]
\centering
\includegraphics[angle=90,width=1.74\columnwidth]{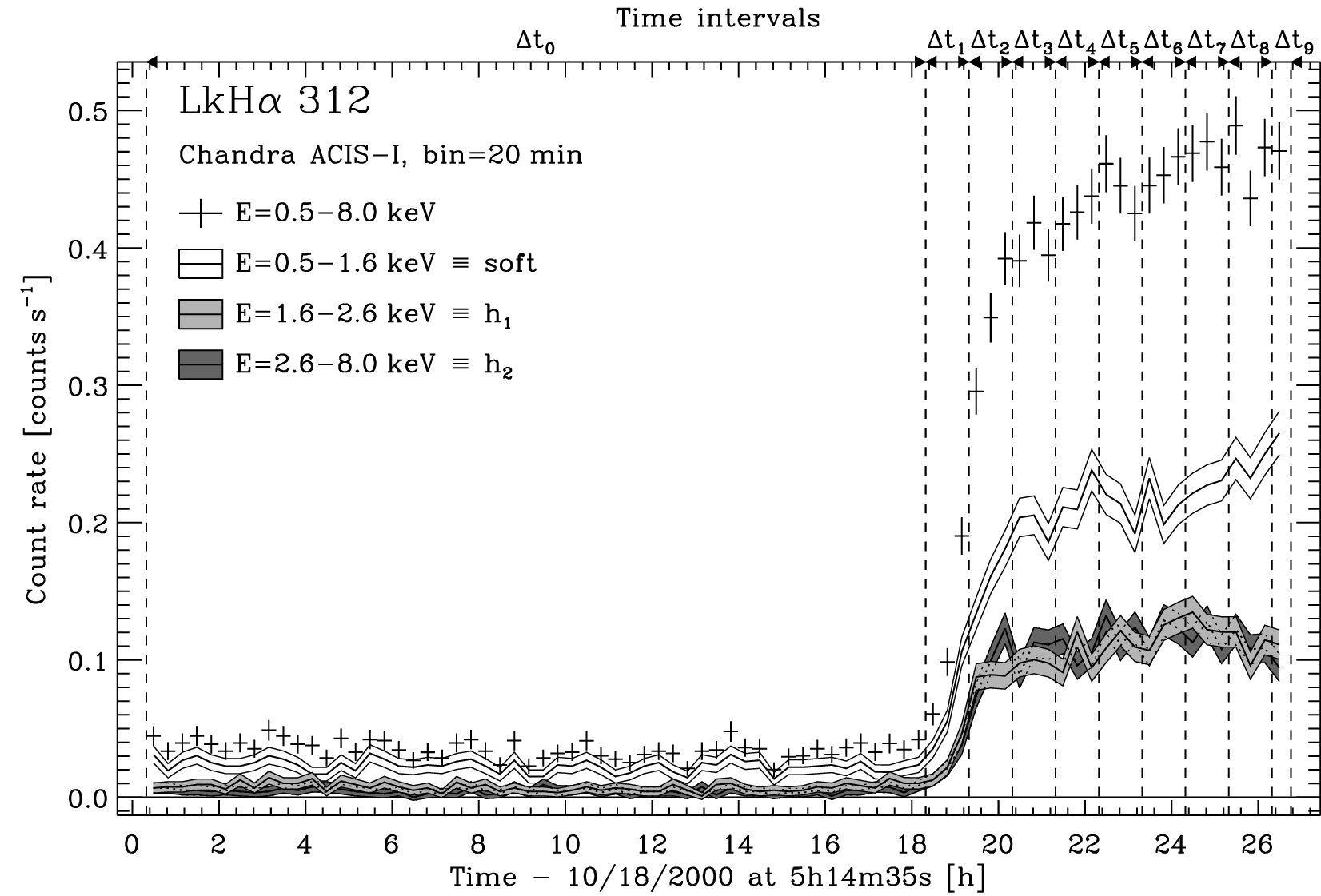}
\caption{Chandra background subtracted X-ray light curves of LkH$\alpha$\,312. The time binning is 20\,min. 
Crosses show the X-ray light curve and one-sigma error bars 
in the energy range 0.5-8\,keV. The stripes show the X-ray light curves 
including one-sigma errors corresponding to the energy ranges 
0.5--1.6\,keV ($soft$ band), 1.6--2.6\,keV ($h_1$ band), and 2.6--8\,keV ($h_2$ band). 
Vertical dashed lines and horizontal arrows define the time intervals, $\Delta$t$_\mathrm{i}$, 
used for the time-dependent spectroscopy. 
The quiescent, fast rise, and gradual phases correspond to the time intervals $\Delta$t$_\mathrm{0}$, 
$\Delta$t$_\mathrm{1}$--$\Delta$t$_\mathrm{2}$, and  $\Delta$t$_\mathrm{3}$--$\Delta$t$_\mathrm{9}$, 
respectively.
}
\label{lightcurve}
\begin{tabular}{@{}cc@{}}
\includegraphics[angle=0,width=\columnwidth]{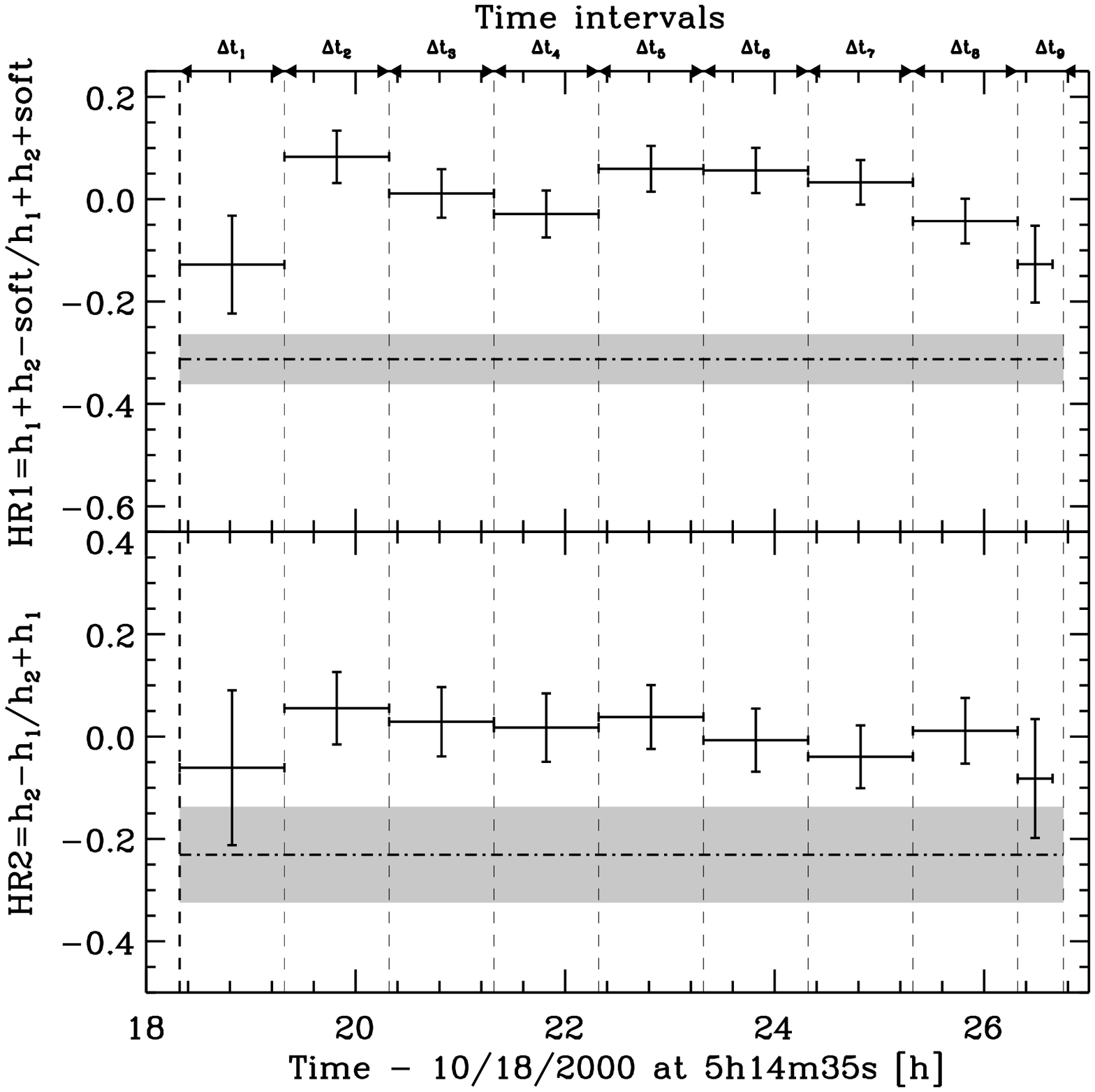}
& 
\includegraphics[angle=0,width=\columnwidth,bb=28 160 566 680,clip]{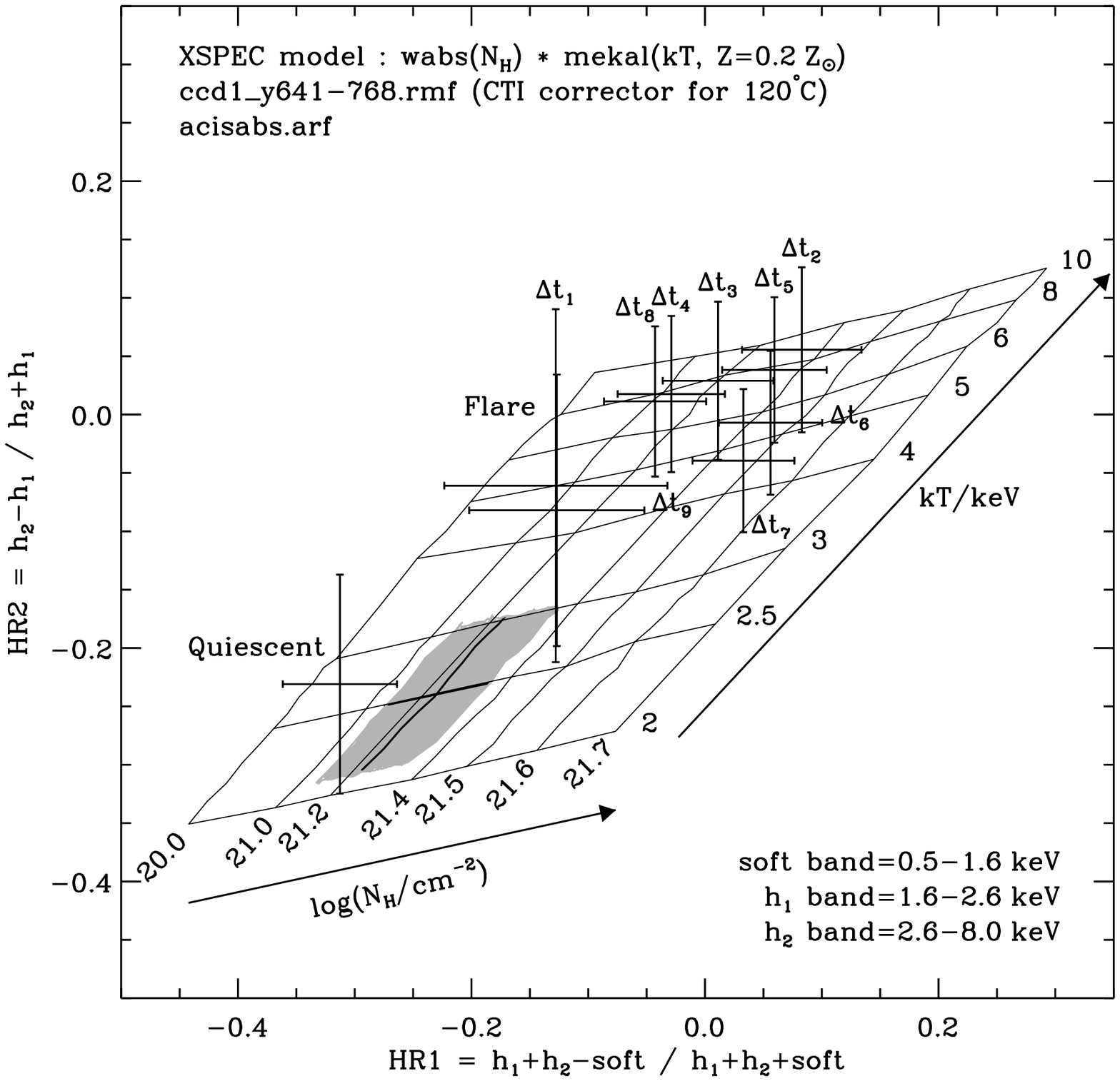}
\\
\end{tabular}
\caption{Left panel~: hardness ratios of LkH$\alpha$\,312 versus time. 
The top and bottom panels show respectively the hardness ratio 
$HR1$ and $HR2$, computed from the counts collected in three energy bands 
($soft$, $h_1$, and $h_2$, see Fig.~\ref{lightcurve}), 
with $1.64\,\sigma$ error bars. 
The grey stripe indicates the average hardness ratio of the quiescent phase.
Right panel~: hardness ratios of LkH$\alpha$\,312 compared to spectral models. 
Iso-energy and iso-absorption lines correspond to absorbed 
optically thin one-temperature plasma spectra model with 0.2 solar abundances, 
and show the correspondence between the hardness ratio pair ($HR1$, $HR2$) and 
the absorption/temperature pair ($N_{\rm H}$, $T$).
The grey area shows the average of two plasma temperatures weighted over the 
emission measure found from spectral fitting of the quiescent phase (see below \S\ref{multi_temperature}, 
Table~\ref{fit_parameters}).
}
\label{hr}
\end{figure*}

We extract source events from the 99\% encircled energy radius, $R(99\%EE)$.
We use the empirical fit given by \cite*{feigelson02b},
$R(99\%EE)=8 + 0.2\,\theta$ (with $\theta$ the off-axis position), 
which leads for $\theta=8.2\arcmin$ to $R(99\%EE)=9.6\arcsec$.
The background is taken from a larger area free of sources.
We use the {\tt CIAO} tool {\tt lightcurve}, which takes into account Good Time Intervals, 
to produce background subtracted light curves 
(corrected from the PSF fraction) with 20\,min bintime, and energy filtering. 

Fig.~\ref{lightcurve} shows the light curve of LkH$\alpha$\,312 in the energy range 0.5--8\,keV.
The X-ray emission was nearly constant during the first 18\,h of the observation, defined as 
the time interval $\Delta$t$_{\rm 0}$, corresponding to the quiescent phase. 
The count rate was then multiplied by a factor of 13 during a fast rise phase ($\sim$2\,h) corresponding 
to $\Delta$t$_{\rm 1}$--$\Delta$t$_{\rm 2}$, and reached a factor of 16 above the quiescent X-ray 
level at the end of a gradual phase ($\sim$6\,h) showing a slower rise and corresponding 
to $\Delta$t$_{\rm 3}$--$\Delta$t$_{\rm 9}$. 
The duration of the fast rise phase is similar to the one observed in other TTS (e.g., \cite{imanishi03}), 
however the gradual phase is really unusual. 
The light curve shape is reminiscent of one of the egress phase of an eclipse; however, 
as we will see below, this event was triggered by plasma heating. 
Therefore LkH$\alpha$\,312 displayed a long duration X-ray flare, 
in contrast to an impulsive flare, during the phase $\Delta$t$_{\rm 1}$--$\Delta$t$_{\rm 9}$.
Some roughly similar events were already observed by {\sl Chandra} in \object{the Orion Nebula Cluster} 
(see for example, \object{JW\,567} and \object{JW\,738} in Fig.~3 of \cite{feigelson02a}), 
or in \object{IC\,348} (\object{CXOPZ\,J034416.4+320955} in Fig.~3 of \cite{preibisch02}) 
but the X-ray counts were lower, and the datastream was shorter.  

The flare count rate was very high, $\sim$0.4--$\sim$0.5\,cts\,s$^{-1}$, allowing time-dependent spectroscopy 
(see \S\ref{spectroscopy}). 
To our knowledge this event has the highest count rate observed so far from a PMS low-mass star with {\sl Chandra}.
Such a high count rate implies a pile-up fraction of $\sim$20\% 
for an on-axis source, however for $\theta=8.2\arcmin$ the PSF spreading reduces 
the pile-up fraction to a value lower than $1\%$ 
(see The Chandra Proposers' Observatory 
Guide\footnote{{\tt http://asc.harvard.edu/udocs/docs/POG/MPOG/\-index.html}}), 
and hence we do not need to worry about this effect in our spectral studies.

\subsection{Hardness ratios}
\label{hardness_ratios}

Before applying automatic fitting procedures to spectra in \S\ref{spectroscopy}, we investigate count ratios 
(`hardness ratio' equivalent to `X-ray colors') variation, to assess absorption and/or temperature 
changes during the flare. We split the 0.5-8.0\,keV energy band into three parts~: the $soft$ band, the $h_1$, 
and $h_2$ hard bands, corresponding to the energy ranges 
0.5--1.6\,keV, 1.6--2.6\,keV, and 2.6--8\,keV, respectively. Fig.~\ref{lightcurve} 
shows the light curves of these three different bands. 
The global shape of these light curves is very similar to the full energy band light curve.  
We compute the following hardness ratios~: $HR1 \equiv (h_1+h_2-soft)/(h_1+h_2+soft)$, 
and $HR2 \equiv (h_2-h_1)/(h_1+h_2)$, with associated errors computed using Gaussian error propagation. 
The left panel of Fig.~\ref{hr} shows the flare hardness ratios versus 
time. They display higher values during the flare phase than during the quiescent phase.

We use rmf and arf files of the quiescent phase of LkH$\alpha$312 (see method to compute 
these files below \S\ref{methodology}) to simulate with {\tt XSPEC} synthetic {\sl Chandra} spectra of absorbed 
({\tt wabs} model with \cite{morrison83} cross-sections) optically thin one-temperature plasma ({\tt mekal} model; 
\cite{kaastra96}) with $Z=0.2\,Z_\odot$ (found from the spectral fitting of the quiescent phase, see below 
\S\ref{multi_temperature} Table~\ref{fit_parameters}), for a grid of absorption ($N_{\rm H}$) and temperature 
($T$) values, and we compute their corresponding hardness ratios. 
The right panel of Fig.~\ref{hr} shows the resulting spectral model grid in the hardness-space with iso-energy 
and iso-absorption lines. Indeed the energy bands were designed to have the best separation of these iso-energy 
and iso-absorption lines, which allow a direct correspondence between the hardness ratio pair 
($HR1$, $HR2$) and the absorption/temperature pair ($N_{\rm H}$, $T$).

The quiescent phase corresponds in this one-temperature plasma model to 
$\log N_{\rm H} \sim 20.7$, and $kT \sim 2.7$\,keV. However as will be shown below 
in \S\ref{spectroscopy} the quiescent phase is in fact better modeled by a two-temperature plasma. 
The plasma temperature found with the hardness ratio method is consistent with the average 
of the plasma temperatures weighted over the emission measures, but the absorption estimate is 
here clearly underestimated, because as it will see below in \S\ref{spectroscopy} (Table~\ref{fit_parameters}) 
the spectral fitting leads to $\log N_{\rm H}=21.2$ with a 90\% confidence 
interval $\log N_{\rm H}$=21.1--21.4. 
We add in Fig.~\ref{hr} our best estimate of the ($N_{\rm H}$, $T$) pair for the quiescent phase 
from our spectral fitting.
We can only conclude with these hardness ratios that during the flare the temperature of the plasma 
increases to $\sim$7\,keV on average. There are no variations of the column density.

\begin{table*}[!t]
\caption{Best fit parameters of the time-dependent spectroscopy with their errors at the 90\% confidence levels ($\Delta\chi^2$=2.71; corresponding to $\sigma=1.64$ for Gaussian statistics). $\eta \equiv \log{(L_\mathrm{X,intr}/L_\mathrm{bol})}$.
}
\label{fit_parameters}
\begin{tabular}{@{}crccccrrrrcc@{}}
\hline
\hline
\noalign{\smallskip}
                        &     &                    &                    & \multicolumn{2}{c}{Temperature} & \multicolumn{2}{c}{Emission Measure}& & & \multicolumn{1}{c}{$\log L_\mathrm{X,intr}$} & $\eta$ \\
\vspace{-0.6cm}\\
                        &     &                    &             &        \multicolumn{2}{c}{\hrulefill} & \multicolumn{2}{c}{\hrulefill}& &  &  \multicolumn{1}{c}{\hrulefill} & \\
$\Delta$t$_\mathrm{i}$  & \multicolumn{1}{c}{$N$} & $N_\mathrm{H,21}$& $Z$   & $kT_\mathrm{low}$ & $kT_\mathrm{high}$ & $EM_\mathrm{low}$ & $EM_\mathrm{high}$   & \multicolumn{1}{c}{$\chi^2_\nu$ ($\nu$)} & \multicolumn{1}{c}{$\cal{Q}$} & 0.5--2/2--8/0.5--8  & \\
                & \multicolumn{1}{c}{[cts]}      & [cm$^{-2}$]   & [$Z_\odot$] & \multicolumn{2}{c}{[keV]} & \multicolumn{2}{c}{[$10^{54}$\,cm$^{-3}$]} & &\multicolumn{1}{c}{[\%]} & \multicolumn{1}{c}{[erg\,s$^{-1}$]}   \\
\noalign{\smallskip}
\hline
\noalign{\smallskip}
0 &  2207 & 1.7$_{-0.5}^{+0.6}$ & 0.2$_{-0.1}^{+0.2}$ & 1.0$_{-0.1}^{+0.2}$ & 3.1$_{-0.6}^{+1.8}$ & 0.2$_{-0.1}^{+0.4}$ & 0.5$_{-0.2}^{+0.1}$ & 1.08 ( 79) & 29 & 30.6 30.4 30.8 & -2.9 \\
\noalign{\smallskip}
1 &   392 & 1.2$_{-1.2}^{+1.3}$ & 0.2$_{-....}^{+1.1}$ & \dotfill$_{}^{}$ & 4.6$_{-1.4}^{+3.1}$ & \dotfill $_{}^{}$ & 1.9$_{-0.6}^{+0.6}$ & 1.04 ( 14) & 41 & 31.0 31.1 31.4 & -2.4 \\
\noalign{\smallskip}
2 &  1195 & 2.4$_{-0.7}^{+0.7}$ & 0.5$_{-....}^{+1.6}$ & \dotfill$_{}^{}$ & 6.9$_{-1.7}^{+3.3}$ & \dotfill $_{}^{}$ & 5.8$_{-1.4}^{+1.1}$ & 1.14 ( 48) & 23 & 31.5 31.8 32.0 & -1.8 \\
\noalign{\smallskip}
3 &  1394 & 1.9$_{-0.6}^{+0.7}$ & 2.1$_{-1.4}^{+3.9}$ & \dotfill$_{}^{}$ & 6.4$_{-1.4}^{+1.8}$ & \dotfill $_{}^{}$ & 5.0$_{-1.8}^{+1.5}$ & 1.15 ( 54) & 21 & 31.6 31.9 32.0 & -1.7 \\
\noalign{\smallskip}
4 &  1493 & 1.7$_{-0.6}^{+0.6}$ & 0.5$_{-0.4}^{+0.6}$ & \dotfill$_{}^{}$ & 5.6$_{-1.1}^{+1.6}$ & \dotfill $_{}^{}$ & 6.9$_{-1.0}^{+1.1}$ & 0.68 ( 57) & 97 & 31.6 31.8 32.0 & -1.7 \\
\noalign{\smallskip}
5 &  1561 & 2.2$_{-0.5}^{+0.6}$ & 0.0$_{-....}^{+0.4}$ & \dotfill$_{}^{}$ & 7.6$_{-1.7}^{+3.2}$ & \dotfill $_{}^{}$ & 8.0$_{-0.8}^{+0.7}$ & 0.85 ( 63) & 80 & 31.6 31.8 32.1 & -1.7 \\
\noalign{\smallskip}
6 &  1590 & 2.8$_{-0.6}^{+0.6}$ & 0.1$_{-....}^{+0.3}$ & \dotfill$_{}^{}$ & 5.3$_{-1.0}^{+1.5}$ & \dotfill $_{}^{}$ & 9.0$_{-1.0}^{+1.1}$ & 0.86 ( 61) & 77 & 31.7 31.8 32.1 & -1.7 \\
\noalign{\smallskip}
7 &  1642 & 2.1$_{-0.5}^{+0.6}$ & 0.5$_{-0.4}^{+0.4}$ & \dotfill$_{}^{}$ & 5.1$_{-0.8}^{+1.2}$ & \dotfill $_{}^{}$ & 7.8$_{-0.9}^{+1.0}$ & 1.06 ( 62) & 36 & 31.7 31.8 32.1 & -1.7 \\
\noalign{\smallskip}
8 &  1641 & 2.1$_{-0.5}^{+0.5}$ & 0.1$_{-....}^{+0.3}$ & \dotfill$_{}^{}$ & 4.8$_{-0.8}^{+1.1}$ & \dotfill $_{}^{}$ & 8.6$_{-1.0}^{+1.1}$ & 1.11 ( 62) & 26 & 31.7 31.8 32.0 & -1.7 \\
\noalign{\smallskip}
9 &   715 & 1.6$_{-0.8}^{+0.9}$ & 0.0$_{-....}^{+0.4}$ & \dotfill$_{}^{}$ & 3.8$_{-0.8}^{+1.2}$ & \dotfill $_{}^{}$ & 9.1$_{-1.7}^{+1.8}$ & 0.56 ( 28) & 97 & 31.7 31.7 32.0 & -1.8 \\
\noalign{\smallskip}
\hline
\noalign{\smallskip}
1--9 & 11327 & 2.6$_{-0.3}^{+0.4}$ & 0.3$_{-0.1}^{+0.1}$ & 0.8$_{-0.4}^{+0.5}$ & 5.1$_{-0.4}^{+0.5}$ & 0.3$_{-0.2}^{+0.3}$ & 7.4$_{-0.3}^{+0.3}$ & 1.09 (227) & 17 & 31.6 31.8 32.0 & -1.7 \\
\noalign{\smallskip}
\hline
\end{tabular}
\end{table*}

\section{Time-dependent spectroscopy}
\label{spectroscopy}

        \subsection{Methodology}
\label{methodology}

The spectral analysis is performed with {\tt XSPEC} 
(version 11.2\footnote{\tt http://lheawww.gsfc.nasa.gov/users/kaa/home.html}). 
Plasma models are convolved with the response matrix file (rmf) describing 
the spectral resolution of the detector as a function of energy, and an 
auxiliary response file (arf) describing the telescope and ACIS detector effective area as 
a function of energy and location in the detector. 
Reduction in exposure times due to telescope vignetting, and possible bad CCD columns 
moving on the source with satellite aspect dithering, is included in the arf file.
We use the rmf files and the quantum efficiency uniformity (QEU) files provided 
with the CTI corrector package. These QEU files are used by {\tt mkarf} 
to generate arf file consistent with our CTI corrected events. 
We note that for LkH$\alpha$\,312 (I1 chip, row 749) far away from the readout node 
(I1 chip, row 1), the CTI correction improves the energy resolution from FWHM$\sim$130\,eV 
to 110\,eV in the soft part of the spectrum, and from FHWM$\sim$260\,eV 
to 190\,eV in the hard part of the spectrum (see \cite{townsley02}).
We apply the {\tt acisabs} model to the arf file to take into 
account the contamination of the ACIS optical blocking filter. 
For our observation 453 days after launch, this correction is negligible above $\sim$1\,keV, 
but leads to a decrease of 20\% of the effective area at 0.5\,keV, which is 
not negligible to determine an accurate value of hydrogen column density on the line 
of sight. Best-fit model parameters are found by comparing models with the pulse height distribution 
of the extracted events using $\chi^2$ minimization. The source spectrum is rebinned with 
{\tt grppha} in the {\tt ftools} package to a minimum of 20 source counts per spectral bin.
The model is improved until obtaining $\chi^2_\nu \sim1$ and {${\cal Q} \ge 5\%$}, 
with $\chi^2_\nu$ the reduced $\chi^2$ (i.e., $\chi^2$ over the degree 
of freedom, $\nu$), and ${\cal Q}$ the probability that one would observe $\chi^2_\nu$, or a larger 
value, if the assumed model is true, and the best-fit model parameters are the true parameter values. 
The parameter error at the 90\% confidence level ($\Delta\chi^2$=2.71; corresponding to $\sigma=1.64$ 
for Gaussian statistics) is computed with the commands 
{\tt error} and {\tt steppar}.

	\subsection{Multi-temperature plasma model}
	\label{multi_temperature}

We first model the plasma in {\tt XSPEC} with an optically thin single-temperature 
plasma model ({\tt mekal} model; \cite{kaastra96}), combined with photoelectric absorption 
using \cite*{morrison83} cross-sections ({\tt wabs} model). We are led to increase the number 
of plasma component to improve the goodness-of-fit.

		\subsubsection{Quiescent emission and flare spectra}
		\label{quiescent_flare_spectra}

\begin{figure}[!t]
\centering
\includegraphics[angle=0,width=8.8cm]{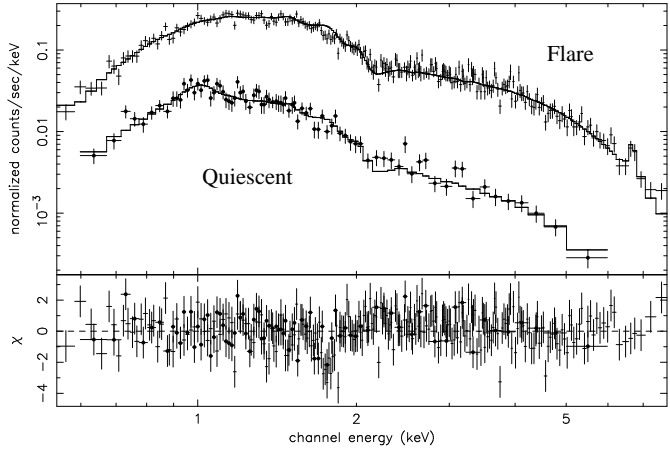}
\caption{Chandra ACIS-I spectra of LkH$\alpha$\,312. The upper panel displays 
separately the ACIS-I data for the quiescent (circle mark) and flare phases. 
The lines are the corresponding models, a two-temperature plasma combined with 
photoelectric absorption (see Table~\ref{fit_parameters} for details). 
The lower panel shows the residual $\chi$ of the modeling. }
\label{spectra}
\end{figure}

Fig.~\ref{spectra} shows the quiescent and the flare spectra for comparison. Both phases require more 
than a one-temperature plasma model to be fitted, and we use a two-temperature model. 
The resulting fit parameters are listed in Table~\ref{fit_parameters}. 
The quiescent spectrum is featureless, no emission lines are visible. 
Approximatively 70\% of the X-ray emission comes from a high temperature plasma at energy $kT=3.1$\,keV. 
The plasma abundance is $Z\sim0.2\,Z_\odot$ (this is the value that was used previously for hardness ratio study in 
\S\ref{hardness_ratios}). 
The average plasma energy, weighted over the emission measure, is $<kT>$=2.5\,keV, 
corresponding to an average temperature of 29\,MK, similar to the temperature found in the previous section with the hardness ratio method.  
The absorption is low with $N_{\rm H}=1.3$--$2.3\,10^{21}$\,cm$^{-2}$ corresponding to $A_{\rm V} \sim $0.7--1.3\,mag 
(\cite{predehl95}).

The intrinsic luminosity in the energy range 0.5--8\,keV is 
$L_{\rm X,intr}=6\,10^{30}$\,erg\,s$^{-1}$. From the stellar radius $R_\star=2.6\,R_\odot$ 
inferred from the H.-R. diagram (see above \S\ref{status}), 
we find a high X-ray stellar flux, $F_{\rm X}=1.4 \, 10^7$ \,erg\,cm$^{-2}$\,s$^{-1}$.
To quantify the X-ray activity we use the logarithm of 
the ratio of the X-ray luminosity corrected from absorption on the bolometric luminosity, 
$\eta \equiv \log (L_{\rm X,intr}/L_{\rm bol})$. (This indicator is independent of the 
source distance.) The quiescent phase displays $\eta=-2.9$. This value is very high compared with
the average value for TTS, $\eta \sim -4$, and implies that the corona of LkH$\alpha$\,312 reached 
the {\sl saturation\,} level.   

The large number of events collected during the flare phase (about 12,000) unveils in the flare spectral 
residual a pseudo-absorption line at 1.785\,keV, which is due to a calibration problem of the quantum 
efficiency just below the Si absorption edge at 1.839\,keV. We suppress the corresponding energy channels 
for the flare phase fitting, which improves greatly the goodness-of-fit. This calibration problem 
is not visible in spectra with only a few thousand counts, and we will neglect it in \S\ref{time_spectro}. 
The flare spectrum is also nearly featureless, with only the He-like line of the Fe\,{\small XXV} 
visible at 6.7\,keV. Approximatively 95\% of the X-ray emission comes from a high energy plasma at 
$kT=5.1$\,keV. We detect a lower energy plasma, which is similar to the one found during the quiescent phase, 
with $kT\sim1$\,keV and $EM\sim0.2\,10^{54}$\,cm$^{-3}$. The flare abundance is also comparable with 
the quiescent one; we detect no metallicity enhancement as found in the bright X-ray flares of 
active stars (e.g., \cite{guedel01}). 
We obtain during the flare phase an accurate measurement of the absorption, 
$N_{\rm H}=2.3$--$3.0\,10^{21}$\,cm$^{-2}$ corresponding to $A_{\rm V} \sim $1.3--1.7\,mag 
(\cite{predehl95}), which is consistent with the optical extinction obtain previously from SED in \S\ref{status}.

Thanks to the high count rate of this flare we can make a 
time-dependent spectroscopy of the rise phase to investigate the physical properties of this event.

\begin{figure}[!t]
\includegraphics[angle=0,width=8.8cm]{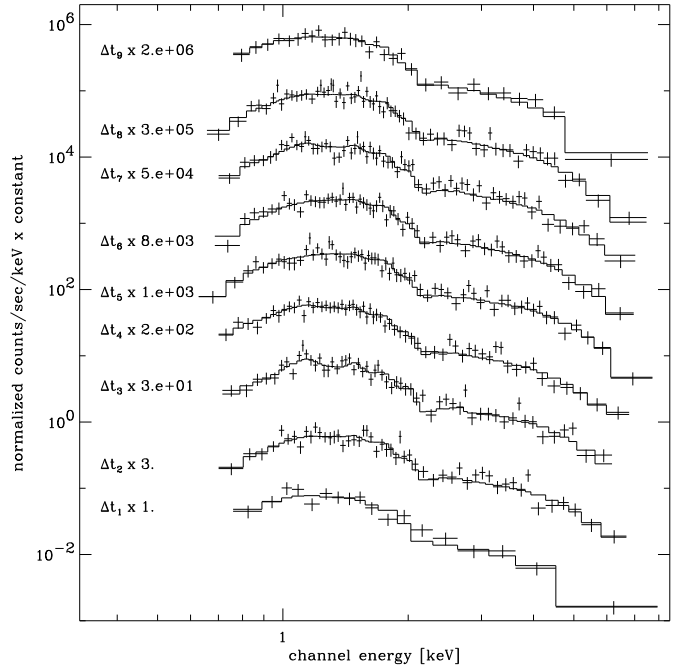}
\caption{Time-dependent flare spectra. Time interval spectra are plotted from bottom to top 
with their one-temperature fit (see Table~\ref{fit_parameters} for details) multiplied by an arbitrary constant.
}
\label{time_flare_spectra}
\end{figure}

\begin{figure}[!t]
\centering
\includegraphics[angle=0,width=8.8cm]{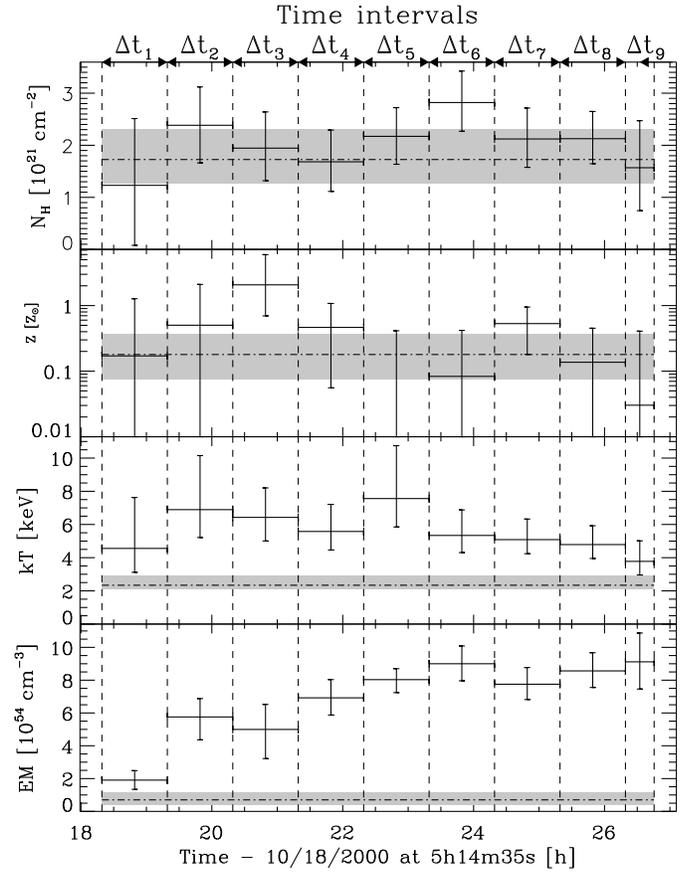}
\caption{Flare parameters versus time (crosses) compared with the quiescent preflare values (grey stripes) 
from values in Table~\ref{fit_parameters}.
The measured values for the quiescent phase are indicated with a dashed-dot line, with a range shown as a grey stripe. 
Time sampling is defined in Fig.~\ref{lightcurve}. 
Since the flare spectra can each be fitted with a single temperature, we take for comparison with the quiescent preflare temperature 
the average of the two plasma temperatures weighted over the emission measures, and for the 
quiescent preflare emission measure the sum of the two emission measures.
}
\label{fit_parameter_plots}
\end{figure}

		\subsubsection{Time-dependent spectroscopy of the flare phase}
		\label{time_spectro}

Fig.~\ref{time_flare_spectra} shows the flare spectra for the time intervals 
$\Delta\mathrm{t}_1$--$\Delta\mathrm{t}_9$, which all have a duration of 1\,h at 
the exception of the shorter interval $\Delta\mathrm{t}_9$ having only about 25\,min.
This bin selection is a compromise between larger bins, which would smooth short term phenomena, 
and shorter bins, which would reduce the number of count. One-hour bins are long enough here to collect 
$\sim$1,500\,counts and hence have enough statistics to derive accurate physical parameters.
The spectra are featureless, there is not enough statistics to detect the 6.7\,keV iron line, 
which becomes appearent only by adding all these time intervals (see Fig.~\ref{spectra}). 
We find a reliable fit only with a high temperature plasma. We tried to add a low temperature 
plasma having the same properties than the one found during the quiescent state, but this did not improve 
significantly the fitting result. The X-ray emission of the flare plasma is hence dominated by this high 
temperature component. However as we saw previously in \S\ref{quiescent_flare_spectra} by adding all the 
flare time intervals, it is possible with enough statistics to constrain this low temperature emission. 
We list all the fit parameters in Table~\ref{fit_parameters}. 

Fig.~\ref{fit_parameter_plots} plots the flare spectral parameters with their errors at the 90\% confidence level 
versus time compared to their quiescent values.
The absorption is consistent during the flare with its quiescent value.
The plasma abundance is also consistent during the flare with its quiescent value 
with maybe the exception of the interval $\Delta\mathrm{t}_3$ where the metallicity is increased by a factor of 10. 
This metallicity enhancement could be explained by photospheric evaporation produced by the flare as seen 
in active stars (e.g., \cite{guedel01}). However assuming that the metallicity keeps its quiescent value during the flare, 
this deviant bin is consistent with being a statistical fluctuation, as in 10 independent measurements one discrepant value 
is expected at the 90\% confidence level.
The emission measure increases until $\Delta\mathrm{t}_6$, and saturates. 

The emission measure behaviour is hence totally different from a rotational modulation 
of a flare decay (\cite{stelzer99}), where the emission measure decreases exponentially.
The plasma temperature is greater than the average quiescent temperature during all the flare, and reaches its maximum 
for $\Delta\mathrm{t}_5$ with $kT=7.6$\,keV. We will consider this maximum as the peak flare in the following section. 
The temperature interval $\Delta\mathrm{t}_5$--$\Delta\mathrm{t}_9$ is reminiscent of the cooling phase of an impulsive flare, 
which displays both an exponential decrease of the temperature and the emission measure, however here no decay 
is visible for the emission measure. 

The luminosity in the energy range 0.5--8\,keV corrected for absorption is 
$L_\mathrm{X,intr}\sim 1 \times 10^{32}$\,erg\,s$^{-1}$ throughout nearly all the flare. 
The X-ray luminosity of the flare peak reaches $\sim$2\% of the stellar bolometric 
luminosity. The total energy released by this flare is $E_\mathrm{X}=7.2\,10^{33}$\,erg in 8.5\,h.

\section{Discussion}
\label{discussion}

The LkH$\alpha$\,312 flare detected here is extraordinary in its
X-ray luminosity and, due to the serendipitous nature of the
off-axis location, in the quality of the X-ray data.  It thus
provides a valuable opportunity to examine the applicability of
established theories for solar-type magnetic reconnection flaring
in the limit of high power and long duration.  Note that we
consider the absorption, consistent with a constant value of 
$N_\mathrm{H} \approx 1.7 \,10^{21}$\,cm$^{-2}$, to be external to the flare
environment and of no interest here.

\subsection{Quiescent emission}
\label{quiescent}

The low X-ray emission prior to a major flares, such as we see in the
first 5 hours of the $Chandra$ observation, is usually considered
to be a `quiescent' plasma component distributed over large
regions as in the solar corona.  But the quiescent emission here
is much hotter and more luminous than the solar corona, with the 
bulk of temperatures ranging from 10--35\,MK
with an average (weighted over the emission measure) of 29\,MK 
(see \S\ref{quiescent_flare_spectra}, Table~\ref{fit_parameters}). 
In contrast, the non-flaring solar corona has
average (weighted over the emission measure) temperature of $\sim2$\,MK (\cite{peres00}). 
The intrinsic quiescent X-ray luminosity in LkH$\alpha$\,312 is 
$6 \, 10^{30}$ erg s$^{-1}$ in the {\sl Chandra} band (0.5--8\,keV) 
and probably $\sim 10^{31}$\,erg\,s$^{-1}$ in a full bolometric X-ray band.  
In contrast, the bolometric X-ray luminosity of the non-flaring solar corona is
only $3 \, 10^{27}$\,erg\,s$^{-1}$ during solar minimum to 
$3 \, 10^{28}$\,erg\,s$^{-1}$ during solar maximum (\cite{peres00}).
Note that the surface area of LkH$\alpha$\,312 is only 6.8 times that of the Sun 
(see \S\ref{quiescent_flare_spectra}), so the X-ray luminosity difference of a factor of order 1,000 
cannot be attributed to a larger surface area.

The higher temperature of the LkH$\alpha$\,312 quiescent X-ray emission could be attributed to
enhanced coronal processes. Theories for solar coronal heating based on the
release of magnetic energy (\cite{parker79}; \cite{parker88}; \cite{priest00}) 
have $kT \simeq f\,B^2/(8 \pi\,n_\mathrm{c})$, where $n_\mathrm{c}$ is the coronal electron density 
and $f$ is the fraction of the coronal magnetic field which is converted into heat.  
A 4-fold increase on average magnetic field strength, for example, could
account for the temperature difference. But it seems unlikely that
the enormous difference in luminosities can be explained in this
fashion.  Since $L_\mathrm{X,quiescent} \propto n_\mathrm{c}^2\, V_\mathrm{c}$ 
where $V_\mathrm{c}$ is the coronal volume, and since $n_\mathrm{c}$ 
cannot be significantly raised without lowering the temperature, 
a $\sim 10^3$ increase in coronal volume would have to be present 
to give the $10^{31}$\,erg\,s$^{-1}$ quiescent luminosity.  
The corona would have to extend $>$10 stellar radii without significant 
density decrease, which seems quite unrealistic. 
Alternatively the X-ray volume filling factor in the corona of LkH$\alpha$\,312 could be 1,000 
times higher than in the solar corona.

The quiescent emission may also arise from the sum of many
eruptive flares while not excluding a contribution from coronal heating processes. 
This is suggested by 
the growing evidence for microflaring as the origin of most of the quiescent X-ray emission 
from magnetically active stars (\cite{guedel03}, and references therein).  At least 10
flares each with $L_\mathrm{X,flare} \le 10^{30}$\,erg\,s$^{-1}$ must continuously be
active over the stellar surface.
The lack of variation in flux or
temperature during the quiescent phase suggests a large number of
small and moderate-sized flares are present, which in turn
suggests a dense coverage of the stellar surface with
high-strength magnetic fields and hence a high surface filling factor. This is consistent with the fact 
that the quiescent X-ray luminosity is found to reach the saturation level 
as shown in \S\ref{quiescent_flare_spectra}.

As the quiescent level appears remarkably stable and bright, our
understanding of the strong microflaring of LkH$\alpha$\,312 could be advanced 
by a follow-up X-ray observation with high spectral resolution. In particular, the density
of the flare plasmas can be estimated using He-like line ratios.
Ne\,{\small IX} at 0.92\,keV is sensitive to densities from
a few $10^{10}$ to a few $10^{12}$\,cm$^{-3}$ (\cite{porquet00}). 
However, the ionization fraction of Ne\,{\small IX} may be low due to the high
temperatures~: only 0.4\% for plasma at $kT$=1\,keV and 0.007\% at 3\,keV 
(\cite{arnaud85}). Mg\,{\small XI} at 1.34\,keV with a
ionization fraction of 6.8\% at $kT$=1\,keV should then be easier to
detect, but it probes only higher density from a few $10^{11}$ to
a few $10^{13}$\,cm$^{-3}$ (\cite{porquet00}). While the current mission gratings with CCD detectors 
are maybe not sufficiently sensitive for this investigation, it will be
possible with the forthcoming X-ray spectrometer aboard ASTRO-E2, 
a next generation instrument using X-ray bolometers.

\subsection{Rise phase}
\label{rise_phase}

The morphology of the LkH$\alpha$\,312 flare rise is distinctively slow ($\tau \simeq 120$\,min) 
in contrast to a typical solar flare rise ($\tau_\odot \simeq 5$\,min), and smooth with no precursors, 
shoulders or secondary events. 
The smoothness may indicate a relatively simple magnetic geometry such as a single bipolar loop 
or bipolar loop arcade without adjacent multipolar magnetic areas.  
The scale length of the loops can be tentatively constrained by assuming
that the magnetic structure confining the X-ray emitting plasma is
stable during the slow and smooth rise. 
Loops suffer magnetohydrodynamical instabilities with characteristic growth
timescale, $\tau$, corresponding to the length scale of the pre-eruption
region, $l$, divided by the ambient Alfv{\`e}n speed (\cite{priest00}), 
$V_\mathrm{A} \propto B\,n_\mathrm{c}^{-1/2}$, with $B$ the average magnetic field strength, and 
$n_\mathrm{c}$ the pre-flare coronal density. We scale the LkH$\alpha$\,312 values to the solar values~: 
$l_\odot=5 \, 10^{10}$\,cm for an average CME (\cite{gilbert01}, and references therein),
$B_\odot \sim 100$\,G (e.g., \cite{parker88}), and $n_\mathrm{c,\odot} \sim 10^9$\,cm$^{-3}$ 
(see \cite{shibata02}, and references therein). We obtain~: 
 
\begin{equation}
l = l_\mathrm{\odot} \, (\tau/\tau_\odot) \, (B/B_\odot) \, (n_\mathrm{c}/n_\mathrm{c,\odot})^{-1/2} \,.
\end{equation}

We point out that this scale length is the scale for the volume of erupted field.  
Generally, it can be larger than the volume from which the X-ray emission arises, 
and in the case of the solar flares the scale length of the erupted region may be 2 to 3
times greater than the scale of the flaring region. Therefore we estimate the following 
upper limit for the height, $h$, of a semi-circular magnetic loop filled by the X-ray emitting plasma~:

\begin{equation}
h_\mathrm{max} = 0.5 \times l/3\,,
\label{hmax0}
\end{equation}

\begin{equation}
h_\mathrm{max} = 2\,10^{11}\,\mathrm{cm} \times (B/100\,\mathrm{G}) \, (n_\mathrm{c}/10^{9}\,\mathrm{cm^3})^{-1/2} \,.
\label{hmax0}
\end{equation}

Replacing the loop height by the loop half-length, $L=\pi\,h/2$, 
leads to the following constraint~:
\begin{equation}
L  \le 10^8\,\mathrm{cm} \times B \, n_\mathrm{c,12}^{-1/2} \,.
\label{h}
\end{equation}

This formula will be used in the next section combined with an estimate of $L$ and $B$, 
to derive a lower limit estimate of the pre-flare coronal density.

\subsection{Peak properties}
\label{peak_properties}

Both X-ray (\cite{favata03}) and radio (\cite{guedel02}) stellar 
peak flare properties show that magnetic loop scaling relationships can be derived 
from the corresponding solar flare properties. We use here the work of Shibata \& Yokoyama
(1999, 2002) which is based on a scaling law relation for the maximum temperature 
of reconnection heated plasma and magnetic field strength derived from MHD simulations 
(Yokoyama \& Shibata 1998, 2001).
We note, however, that this work is based on MHD simulations where the effect 
of the radiative cooling can be neglected, so scaling laws which include the cooling 
switching from conductive to radiative as time proceeds (\cite{cargill95}) have not yet 
been numerically tested. 

The theory of Shibata \& Yokoyama involves a roughly semi-circular magnetic
loop with half-length $L$, magnetic field $B$, 
and electron density after chromospheric evaporation $n_\mathrm{e}$. 
These parameters are derived from the peak X-ray temperature 
and emission measure together with prior knowledge of the pre-flare 
coronal electron density $n_\mathrm{c}$, using the following scaling formulae where $kT$ is expressed in keV~:
\begin{equation}
L    =  2\,10^{11}\,\mathrm{cm} \times EM_{54}^{3/5} \, kT^{-8/5} \, n_\mathrm{c,12}^{-2/5}\,\rm,
\label{L}
\end{equation}
\begin{equation}
B    =  32\,\mathrm{G}  \times EM_{54}^{-1/5} \, kT^{17/10} \, n_\mathrm{c,12}^{3/10}\,\rm,
\label{B}
\end{equation}
\begin{equation}
n_\mathrm{e} =  10^{10}\,\mathrm{cm}^{-3} \times EM_{54}^{-2/5} \, kT^{12/5} \, n_\mathrm{c,12}^{3/5}\,\rm.
\label{n_e}
\end{equation}

The loop parameters are hence function of $n_\mathrm{c,12}$, which can be constrained.
First, for consistency with the chrosmospheric evaporation assumption, the flare coronal density must be
lower than the pre-flare coronal density, which leads using (\ref{n_e}) to the following upper limit estimate for 
$n_\mathrm{c,12}$~:

\begin{equation}
n_\mathrm{c,12} \le 10^{-5} \, EM_{54}^{-1} \, kT^{\,6} \,\rm.
\end{equation}

Second, from (\ref{h}) and using (\ref{L})--(\ref{B}) leads to a lower limit on $n_\mathrm{c,12}$~:
\begin{equation}
n_\mathrm{c,12} \ge 10^9\, EM_{54}^4 \, kT^{-33/2} \,.
\end{equation}
As the LkH$\alpha$\,312 peak flare values ($EM_{54}=8.0$ and $kT=7.6$\,keV) are consistent with the empirical correlation 
(see Fig.~1 of \cite{shibata02}),
no special scaling laws are required for this flare. Applying the previous set of formulae, we find~:

\begin{equation}
0.01 \le n_\mathrm{c,12} \le 0.24\,\rm,
\end{equation}
\begin{equation}
L=2.3 \, 10^{10}\,\mathrm{cm} \times n_\mathrm{c,12}^{-2/5} \,\rm,
\end{equation}
\begin{equation}
B=670\,\mathrm{G}\times n_\mathrm{c,12}^{3/10} \,\rm,
\label{magnetic_field}
\end{equation}
\begin{equation}
n_\mathrm{e} =  6.4 \, 10^{11}\,\mathrm{cm}^{-3}\times n_\mathrm{c,12}^{3/5}\,.
\label{ne}
\end{equation}

Numerically, we find that the loop half-length ranges from 4.8--16$\,10^{10}$\,cm (0.3--0.9\,R$_\star)$,
the loop height ranges from 3.1--10$\,10^{10}$\,cm (0.2--0.5\,R$_\star$, i.e., 0.5--1.3\,R$_\odot$),
the magnetic field from 430--180\,G, and the flare loop densities from 2.4--0.4$\times 10^{11}$\,cm$^{-3}$.

\subsection{Gradual phase and decay phase}

The gradual phase, which has a slower rise and follows the fast rise phase, is really unusual. 
It may be explained by the rapid fall of the stellar magnetic field with height.
Indeed in this situation when the magnetic loop system rises, the reconnection site is  
continually moving up into a weaker field region, which produces the decline of the rate 
of magnetic energy release. 

Much of solar and stellar flare loop modelling is based on the
decay phase of the flare when the loop plasma is cooling 
(see review by \cite{reale02}; e.g., \cite{reale97}). 
Unfortunately, only the beginning of this phase may have been observed in LkH$\alpha$\,312 
in the second part of the gradual phase where we find a decrease 
of the plasma temperature from 88\,MK (7.6\,keV) to 44\,MK (3.8\,keV) 
over 5 hours using one-hour averaged spectra (Figure~\ref{fit_parameter_plots}).  
However during this period, the broad-band luminosity remained constant, 
which is rather unusual for a cooling period, so that the emission measure
(density and/or volume) increased somewhat as the plasma cooled, or short heating phases may occur 
which are smoothed by the averaged spectra.

\section{Conclusions}
\label{conclusion}

So far, LkH$\alpha$\,312 was considered from $\sim$30-yr old H$\alpha$ prism surveys as an unremarkable,
weak-activity star at the periphery of M\,78, until {\sl Chandra} unveiled its extraordinary activity in X-rays.
Our study based on the available optical and IR data shows that it is likely a $< 1$ Myr-old T Tauri star
similar to the Sun ($M_\star \sim 0.7$--0.75\,M$_\odot, R_\star \sim 2.6$\,R$_\odot$), about as luminous
($L_\mathrm{bol} \sim 1.5$\,L$_\odot$), and surrounded by a hollow circumstellar disk. It is therefore
perhaps a transition object between an active accreting `classical' T Tauri star, and a genuine disk-free 
`weak-line' T Tauri star. 

In contrast with these fairly unoriginal properties of LkH$\alpha$\,312, our {\sl Chandra} observations revealed a particularly long and intense X-ray flare, in addition to a bright 
quiescent emission. Because of the intensity of the flare ($\sim 0.5$\,{\sl Chandra} cts\,s$^{-1}$)
and its favorable far off-axis position on the ACIS-I detector which prevented pile-up, the quiescent state
as well as the rise phase of the flare could be studied at an unprecedented level of detail,
yielding spectroscopy resolved on a time scale of down to 1\,h out of a 26\,h-long observation.
The flare itself is quite remarkable, with its high X-ray luminosity (comparable to that of only
a handful of TTS so far) and slow rise, but so is the level of the quiescent emission, which is saturated.
The high intensity of the quiescent level was already perceptible in the {\sl ROSAT/HRI} observation of 1997,
so it is likely not a brief, transient phenomenon.

Quantitatively, the X-ray properties
derived from our {\sl Chandra} observations are as follows~: (i) the quiescent emission of LkH$\alpha$\,312 is stable over 18\,h of observation and saturated
at a high level of activity ($L_\mathrm{X}/L_\mathrm{bol} \sim 10^{-3}$, with $L_\mathrm{X} \sim 6 \times 10^{30}$\,erg\,s$^{-1}$); 
(ii) LkH$\alpha$\,312 also displayed an unusually long
and intense X-ray flare, showing a long rise phase extending over several hours.
This flare was bright, reaching $L_\mathrm{X} \sim 10^{32}$\,erg\,s$^{-1}$, i.e., $\sim$2\%
of the bolometric luminosity. Perhaps more remarkably, the peak temperature was quite high ($kT \sim 7$\,keV). 
(iii) In terms of solar-type magnetic confinement of the plasma,
the application of the scaling laws found by Shibata \& Yokoyama shows that the flare
corresponds to a fairly large volume (maximum height of about 1\,R$_\odot$), and
typical solar coronal densities ($\sim 10^{11}\,{\rm cm}^{-3}$).

The problem is to understand the reasons for these rather extreme X-ray properties, both in the quiescent state and in the flaring state. The first, ``standard''
way is to address the problem in the framework of solar physics. This is how we derived the flare
properties (size and density of the confining magnetic loop, strength of the magnetic field). As discussed
in \S\ref{quiescent}, one could invoke enhanced coronal processes and/or microflaring, 1,000 times more intense than the quiet Sun, or a volume filling factor in the LkH$\alpha$\,312 corona 1,000 times larger than in the solar corona, but the physical reasons for such an
enhancement are unclear. This is in fact a general problem for pre-main sequence stars~: the mechanism for magnetic field generation leading to the high level of X-ray activity is still not well
understood. For instance,
the lack of an X-ray luminosity/rotation correlation on large samples like in Orion is perhaps due to the fact that an $\alpha^2$ (turbulent)
dynamo operates at the young TTS stage, as opposed to the standard rotation-convection induced $\alpha-\omega$
dynamo which holds in late-type main-sequence stars (\cite{feigelson03}).

An alternative way is to consider that the enhanced X-ray activity of  LkH$\alpha$\,312, and perhaps more generally
of at least some other, very X-ray luminous TTS, is triggered by an external cause such as a close,  faint companion like
a brown dwarf or even a planet. Indeed, as shown by \cite*{cuntz00}, {\it planets} can cause enhanced magnetic activity if they are
sufficiently massive and close to the parent star, like 51\,Peg-like planet-bearing stars. The general idea
is that tidal effects can enhance the dynamo by altering the outer convective zone, and/or, if the planet is magnetized, the star and the planet can
interact magnetically via shearing and reconnections, for instance as in the star-disk magnetic interaction mechanism proposed
by \cite*{montmerle00} to explain the triple X-ray flare of the YLW15 protostar. In such a framework, the rotation information would be important~: although a slow rotation would not be conclusive, a fast rotation (periods of days or less) could be indicative of a tidally locked, RS CVn-like binary system. We note that the circumstellar disk of LkH$\alpha$\,312 is hollow, with no accretion taking place,
which is indeed consistent with the possible presence of a planet clearing the inner disk (e.g., \cite{nelson03}).

Both ground and space follow-up observations will be needed to investigate
the coronal properties of LkH$\alpha$\,312 as a genuine TTS and its influence on its circumstellar material~:
medium resolution optical spectroscopy, to confirm its youthfulness with Li\,{\small I}~6707
absorption line measurement, to check that the H$\alpha$ emission has not changed, and to estimate its projected equatorial velocity;  to determine its
rotational period from the photometric modulation of starspots; high-resolution IR spectroscopy to detect the possible presence of a very low mass companion; 
near-IR high resolution imagery to study a possible transition circumstellar disk; and of course high resolution X-ray spectroscopy to probe the density of its coronal plasma.

\begin{acknowledgements}
We would like to thank the anonymous referee for comments and suggestions,
Leisa Townsley and Pat Broos to have introduced N.G. to the CTI corrector and the TARA package, 
Pr. Gordon Garmire for his hospitality during N.G. stay in the Davey Laboratory, and Pr. Eric Priest for his suggestions. 
N.G., E.D.F., and T.G.F.\ acknowledge support from MPE stipendium and CNRS postdoc, SAO grant GO0-1043X, and from NASA grants NAG5-8228, and NAG5-10977.
This publication makes use of the ROSAT Data Archive of the Max-Planck-Institut f{\"u}r extraterrestrische 
Physik (MPE) at Garching, Germany, and data products from the Two Micron All Sky Survey, 
which is a joint project of the University of Massachusetts and the Infrared Processing 
and Analysis Center/California Institute of Technology, funded by the National 
Aeronautics and Space Administration and the National Science Foundation. 
\end{acknowledgements}

\end{document}